 \documentclass[final,5p,times,twocolumn]{elsarticle}

\usepackage{graphicx}

\usepackage{amssymb}

\biboptions{comma,round}

\newcommand\ltt{$L_{3,2}$ }
\newcommand\altt{$L_3$ }

\newcommand\etal{{\it et\ al\ }}

\journal{Carbon}

\begin{document}

\begin{frontmatter}



\title{Electronic and magnetic structure of carbon nanotubes using x-ray absorption and magnetic circular dichroism spectroscopy}


\author[kist,pls]{Sanjeev {Gautam}\corref{cor1}} \ead{sgautam71@kist.re.kr}
\author[esrf]{P. {Thakur}\fnref{dls}}
\author[kaist]{S. {Augustine}}
\author[kaist]{J.K. {Kang}}
\author[pls]{J. -Y. {Kim}}
\author[esrf]{N.B. {Brookes}}
\author[nsc]{K. {Asokan}}
\author[kist]{Keun Hwa {Chae}}
\cortext[cor1]{corresponding author. Tel./fax: +82 54 279 1503/1599}
\address[kist]{Nano Analysis Center, Korea Institute of Science and Technology, Seoul 136 791, Republic of Korea}
\address[pls]{Pohang Light Source, San31, Hyojadong, Namgu, Pohang 790 784, Republic of Korea}
\address[esrf]{European Synchrotron Radiation Facility, BP 220, F-38043 Grenoble Cedex, France}
\address[kaist]{Material Science and Engineering, Korea Advanced Institute of Science and Technology, Daejeon 305 701, Republic of Korea}
\address[nsc]{Inter-University Accelerator Center, New Delhi 110 067, India}
\fntext[dls]{Presently at: Diamond Light Source Ltd., UK}

\begin{abstract}
Carbon nanotubes are a fraction of the size of transistors used in today's best microchips, as it could reduce power demands and heating in next electronics revolution. Present study investigates the electronic and magnetic structure  of multi walled carbon nanotubes (MWCNT) synthesized by chemical vapor deposition technique using near edge x-ray absorption spectroscopy (NEXAFS) measurement at C K-edge and x-ray magnetic circular dichroism (XMCD) at Co and Fe \ltt -edges. NEXAFS at C K-edge shows significant $\pi$-bonding, and Fe(Co) L-edge proves the presence of Co$^{2+}$ and Fe$^{2+}$ in octahedral symmetry, and embedded in C-matrix of MWCNT. Element specific hystersis loops and XMCD spectra clearly shows that these MWCNTs exhibits room temperature ferromagnetism. These measurements elucidated the electronic structure of CNTs and presence of magnetic interactions at room temperature.
\end{abstract}


\end{frontmatter}

\section{INTRODUCTION}\label{introduction}
Carbon nanotubes (CNT) possess unique mechanical and electronic properties suitable for fabricating the nano-scale building blocks in nanodevices. Such nanoscale magnetic materials are useful in spin-dependant electronic devices and magnetic data storage. Inducing magnetism in carbon structures could have far reaching consequences for future carbon-based electronics. Magnetism can be induced in carbon structures by defects such as vacancies or impurities. In the absence of defects, all of the bonding electrons are paired in $\pi$ bonds. However, defects which delocalize one of the pair bonds induce excess spin polarization. This can lead to ferromagnetism (FM) when the defects are sufficiently dense. In graphite, impurity-induced FM has been demonstrated and studied theoretically \cite{r1,r2,r3,r4,r5,r6,r7,r8,r9,r10}. Since magnetic materials have been traditionally associated with the elements containing partially filled 3$d$ or 4$f$ subshells which possess the strongly localized nature and high degree of degeneracy. FM appearing in carbon-based materials has attracted much attention because carbon contains only $sp$ electrons and hence it makes carbon to be considered as a candidate for nanoscale magnetic devices and for spintronics. 
\begin{figure}[tbh!]\centering
  \includegraphics[width=6.5cm,angle=90]{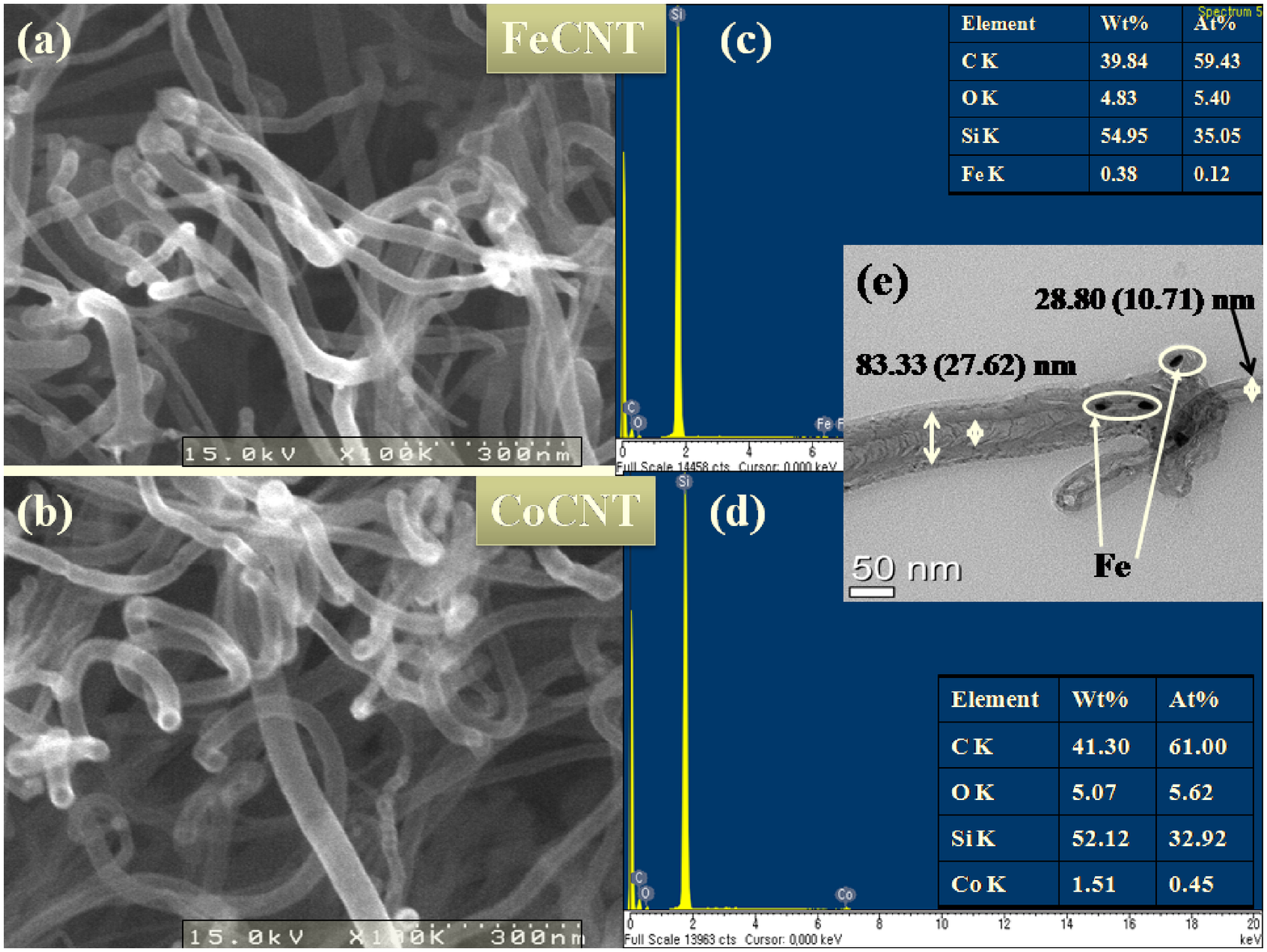}
  \caption{(Color online) SEM, EDAX and TEM micrographs for CNTs (a) Fe-CNT (b) Co-CNT (c) EDAX for FeCNT (d) EDAX for CoCNT (O and Si contents are observed from -the substrate used) (e) TEM micrograph for FeCNT, showing Fe-nanoparticles on the CNT's walls.}\label{fig1}
\end{figure}
However, the origin of magnetism in carbon materials is still controversial. Both the edge states appearing around zigzag shaped edges \cite{r10,r11a,r11b} and the defects (vacancy or hydrogen chemisorption defects) \cite{r12,r13}  are theoretically predicted to result in magnetism. The two primary contributors to this magnetism are H-complex and adatom magnetism. In H-vacancy complex magnetism, hydrogen bonds to a carbon atom near a carbon vacancy creating a new $\sigma$ bond, thereby inducing spin polarization in an unpaired $\pi$-electron \cite{r13,r14}. If the CNT is multi walled, then a transition metal may replace the hydrogen, connecting the walls of CNT. The increased interlayer separation in ordered graphite provides an appropriate length scale for TM-ion adsorption and the formation of carbon interstitials and Frenkel pairs. On the other hand, a defect caused by a carbon adatom would have the same effect as in the TM-complex of delocalizing the nearby $\pi$-electrons by becoming saturated by TM-ion, resulting in adatom ferromagnetism. However, at high defect densities a significant number of adatoms will likely form covalent bonds in nearby vacancies \cite{r14,r5}. It was also postulated that oxygen and nitrogen incorporation can lead to ferromagnetism in carbon structures \cite{r16}. Following these predictions, defect-induced magnetism was found in experiments on graphite \cite{r17} and Films of highly-ordered pyrolytic graphite (HOPG) \cite{r4}.

Near edge x-ray absorption spectroscopy (NEXAFS) is a powerful tool revealing the features in the unoccupied density of states (DOS). It has been widely used to study materials made of carbon atoms in its three possible hybridizations: $sp$, $sp^2$, and $sp^3$. The electronic structure of graphite has been widely studied experimentally and theoretically and the $\sigma^*$ and $\pi^*$ thresholds have been examined by various authors \cite{r16,r17,r18}. Combination of NEXAFS and x-ray magnetic dichroism (XMCD) is commonly used to understand the electronic structure of magnetic materials. In this paper, we report on the electronic structure and room temperature ferromagnetism in MWCNTs synthesized by chemical vapor deposition technique using Fe and Co as catalyst. Techniques such as scanning electron microscopy (SEM), tunneling electron microscopy (TEM), Energy Dispersive Analysis of X-ray (EDAX), among others have been able to identify the characteristic features of the electronic structure of MWCNTs. Using NEAXFS and XMCD as a bulk probing technique, we show here that Fe and Co exist in the ionic form on the walls of MWCNTs and shows ferromagnetic behavior at room temperature.
\section{EXPERIMENTAL PROCEDURE}\label{exp}
Various thickness of Fe ($\sim$ 3.5 to 15 nm), and Co ($\sim$3.5 to 15 nm) layers were deposited onto SiO$_2$/Si substrate via RF sputtering with power of 100 W at a substrate temperature of 2000 $^o$C. Then, the substrate was moved to the Microwave Plasma enhanced Chemical Vapour Deposition (MPCVD) chamber \cite{r19} which was evacuated to about $0.1$ Torr. Next, the substrate was heated to 620 $^o$C in vacuum by a halogen lamp. After that, the N$_2$ gas (100 sccm) was flown into the chamber and the substrate was treated by the N$_2$ plasma created using a microwave power of 700 Watt for 45 seconds. Chamber pressure at the time of Plasma Etching was {\bf $18$ Torr}. Well aligned Carbon Nanotubes were grown on Fe, and Co nanoparticles formed by plasma etching by introducing 15 sccm of CH$_4$ along with $85$ sccm of N$_2$ gas and by using a plasma power of 700 Watt. CNT were grown for about 8 minutes. Transmission electron microscopy (TEM) images show that the tubes are polycrystalline and highly disordered. The samples were annealed in a tube furnace containing equal parts Ar and H$_2$ up to 800 $^o$C. The 2A MS (EPU6) beamline of the PLS-I (South Korea) was used to obtain C K-edge spectra in total-electron yield (TEY) and total fluorescence yield (TFY) mode. XMCD experiment is performed at ID08 beamline of ESRF, France. Above measurements were done under the base pressure better than $3\times10^{-10}$ Torr and resolution $0.01$ eV at $10$ K and $300$ K. All the observed spectra were normalized to incident photon flux impinge into the sample monitored by highly transparent gold mesh.
\section{RESULTS AND DISCUSSIONS}\label{results}
\begin{figure}[tbh!]\centering
  \includegraphics[width=8cm]{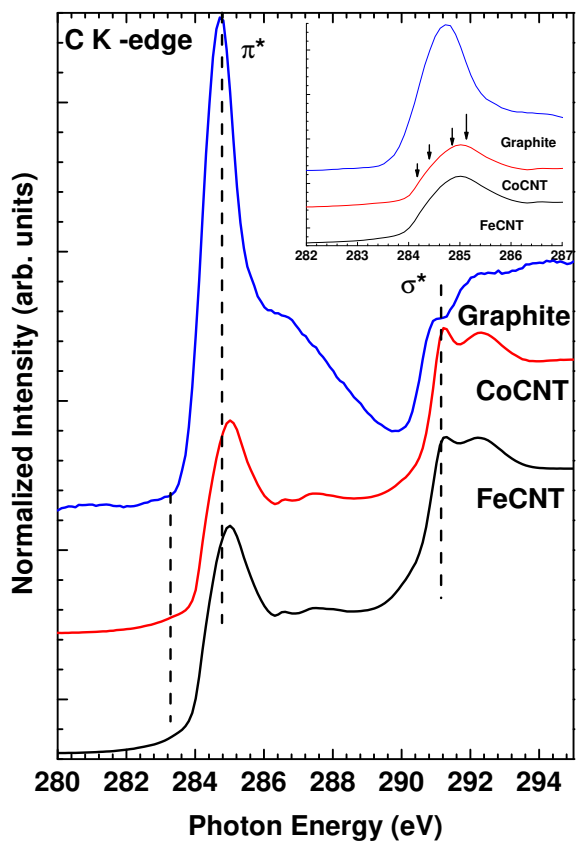}
  \caption{(Color online) C 1$s$ XAS spectra for FeCNT and CoCNTs plotted with the HOPG reference spectra. Inset shows the fine spectral feature at $\pi$* resonance.
}\label{fig2}
\end{figure}

Figure \ref{fig1} (a -e) shows the SEM and TEM micrographs with EDAX data for FeCNT and CoCNTs. It is observed that FeCNTs are $70-80$ nm and CoCNTs are $90-100$ nm in diameter (Fig. \ref{fig1}(a-b)). Fe and Co ions are observed on the walls of MWCNTs as seen in the Fig. \ref{fig1}(e). Fig. \ref{fig1} (c-d) shows EDAX plots for the elemental composition of the MWCNTs. It is found that Fe(Co) contents are $0.38(1.51)$ in weight\%, while Si and O are observed from the substrate (SiO$_2$). Studying the C 1$s$ response in XAS is very useful way to prove the conduction band of the system and this feature has been analyzed for various carbonaceous materials along the years. The overall XAS spectra in the C 1s response are depicted in Fig. \ref{fig2}.

The lines depicted from top to bottom are high-resolution XAS spectra recorded on C 1s edge of highly oriented pyrolitic graphite (HOPG), CoCNT and FeCNT. The topmost high resolution XAS spectrum of graphite shows the well known $\pi$* resonance at 285 eV and the $\sigma$* threshold at $291.2$ eV \cite{r20}. The XAS spectrum of the HOPG was recorded at 45$^o$  to normal incidence. The overall shape of the C 1s core level excitation spectrum of MWCNT is very similar to that of a broadened spectrum of graphite. A comparison to directional-dependent C 1$s$ excitation measurements of graphite \cite{r17,r21} reveals that the C 1$s$ spectrum of the nanotubes is reminiscent of an isotropic average of the in-plane and out of-plane graphite spectra, which is consistent with the isotropic nature of bulk nanotube material \cite{r22}. The FeCNT and CoCNT C K-edge spectra is wider than graphite and seems to have fine spectral features as observed by Kramberger \etal The MWCNT's $\pi$* peak is also upshifted by about 1 eV due to the C 1s core hole effects \cite{r23}, which indicates its excitonic nature observed in SWCNTs \cite{r17}. The core hole e?ect in the broad C 1$s$ $\pi$* resonance resembles bulk $sp^2$ carbon, whereas the small core hole effects in the resonance due to vHs (van Hove singularties) are attributed to molecular excitations \cite{r24}.

\begin{figure}[tbh!]\centering
  \includegraphics[width=8cm]{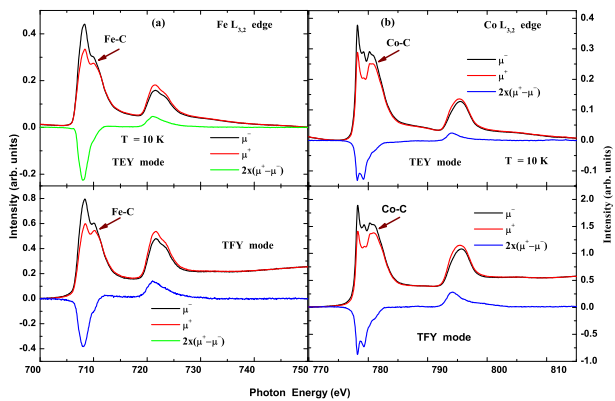}\\[-0.5cm]
  \caption{(Color online) XMCD spectra for MWCNTs at 10 K. (a) Fe \ltt -edge for FeCNT in TEY(surface sensitive) and TFY (bulk sensitive) mode. (b) Co \ltt edge spectra for CoCNTs in TEY (surface sensitive)and TFY (bulk sensitive) mode.}\label{fig3}
\end{figure}

The XMCD signals (Fig.\ref{fig3}) are the difference between XAS spectra recorded for parallel $\mu^+$ and antiparallel $\mu^-$ alignments of the photon helicity with the applied field at 10 K. As evident a clear XMCD ($\mu^+-\mu^-$) with a negative sign confirms the presence of Co$^{2+}$ and Fe$^{2+}$ ions in CoCNT and FeCNT respectively.

\begin{figure}[tbh!]\centering
  \includegraphics[width=8cm]{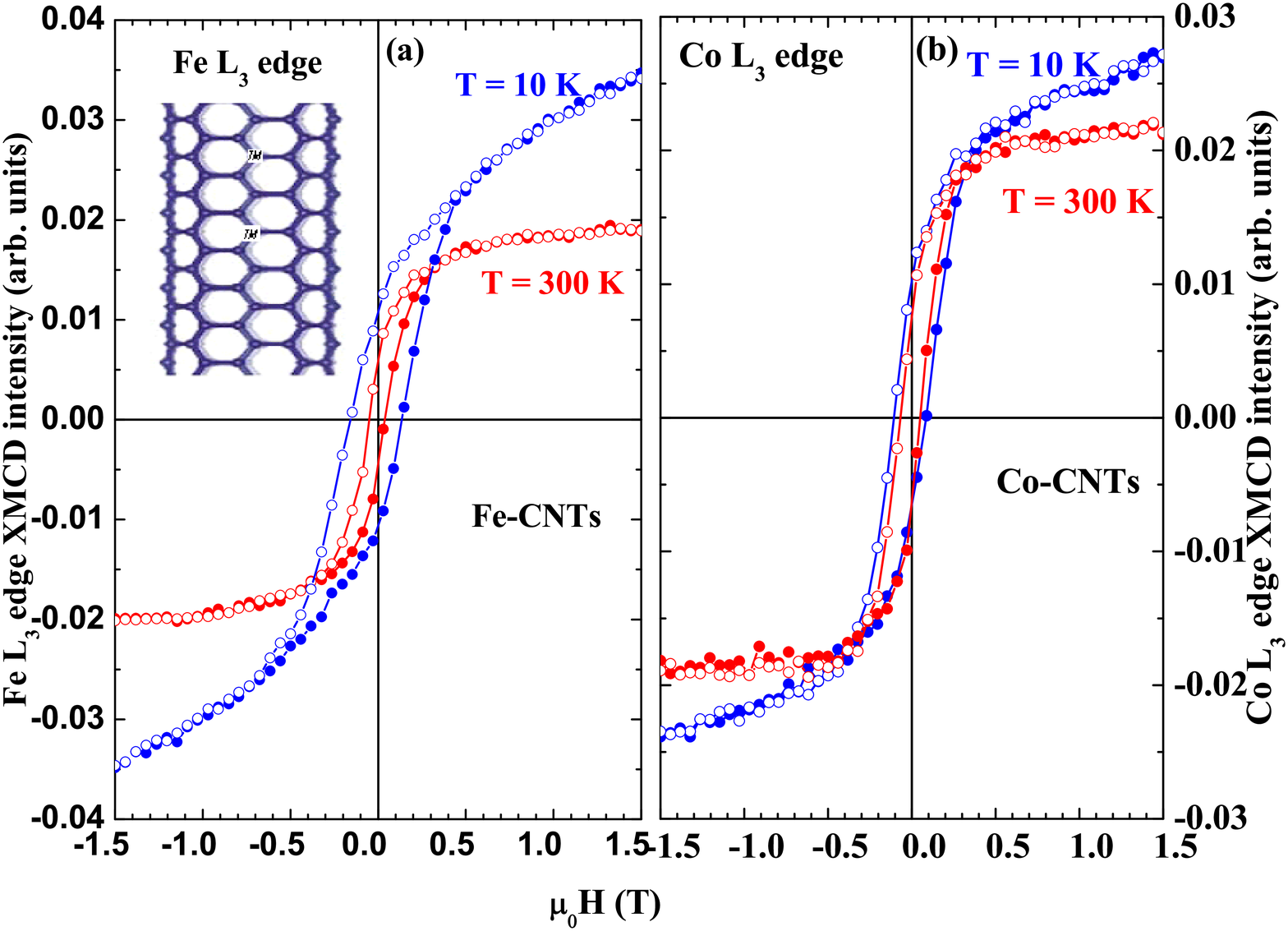}\\[-0.5cm]
  \caption{(Color online) Hysteresis curves at the main peaks of Fe and Co edge at 10 K and 300K. (a) at Fe \altt-edge for FeCNT. (b) at Co \altt-edge for CoCNT. Here both polarization of the photons are used to extract the curves. As evident, at low magnetic field ($< 1$T) an observable ferromagnetic Fe (Co) spins are clearly seen.}\label{fig4}
\end{figure}

Fig. \ref{fig3} show the XMCD spectra for Fe and Co CNTs at 10K in TEY and TFY mode showing that the observed XAS and XMCD spectra is a bulk property and not just the surface. A small change in the XMCD intensity is observed at 300 K (spectra not shown). As evident from the Figures \ref{fig3} (a) and (b), a clear hybridization between Fe(Co) 3$d$ and C 2$p$ orbitals is observed and the Co(Fe) atoms/ions are embeded in the structure of MWCNT as shown in the inset of Fig. \ref{fig4} (a). The same is also observed in HRTEM micrographs (Fig. \ref{fig1}(e)).
Figure \ref{fig4} (a) and (b) displays the hysteresis curves at the main peaks of Fe and Co \ltt -edge at 10 K and 300 K. It is observed that at low magnetic field ($<$ 1T), a clear signature of ferromagnetic components is seen, which is associated with Fe(Co) spins in C-matrix in MWCNTs at room temperature. The observed RTFM is likely originated from Fe(Co) nanoparticles which are on the walls of MWCNT or on the substrate as catalyst. At low temperature, some part of the spin are shows antiferromagnetic/ paramagnetic behavior, which may be due to the presence of small clusters of Fe(Co) in the system.
\section{CONCLUSIONS}\label{conc}
Multiwall Carbon nanotubes (MWCNTs) are successfully synthesized by chemical vapor deposition (CVD) technique. X-ray absorption spectroscopy (XAS) measurement at C K-edge indicates enriched pi-resonance in these MWCNTs. X-ray magnetic circular dichroism (XMCD) at Co and Fe \ltt -edges shows that these MWCNTs exhibit room temperature ferromagnetism. X-ray absorption spectroscopy (XAS) at C K-edge shows significant pi-bonding and Fe(Co) L-edge proves the presence of Co$^{2+}$ and Fe$^{2+}$ in octahedral symmetry.
\section*{ACKNOWLEDGMENT}
This research is supported by the Korea Institute of Science and Technology (KIST, Grant No. 2V01680). Pohang Light Source is supported by MEST, Korea. 





\bibliographystyle{model3-num-names}
\bibliography{cnt-xmcd}







\end{document}